\documentclass[oneside]{article}
\usepackage{geometry,fancyhdr,graphicx}
\usepackage[colorlinks=true, linkcolor=blue, citecolor=blue,urlcolor=blue, breaklinks=true]{hyperref}
\fancypagestyle{plain}{%
  \fancyhf{} % clear all header and footer fields
  \lhead{\raisebox{-2ex}[0pt][0pt]{% includes a vertical correction for the title page header
    \IfFileExists{BiomathShell.pdf}{%
      \includegraphics[height=6ex]{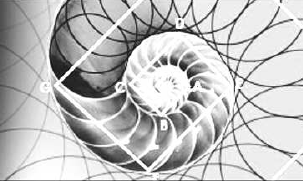}}{%
      \fbox{\parbox[b][1.5em][c]{5ex}{\tiny Missing picture file!}}%
    }%
    \hspace{1.5ex}%
    \shortstack[c]{\sffamily\large Symposium on Biomathematics and Ecology \\[1pt] \sffamily\large Education and Research}%
  }}
  \rhead{\raisebox{-2ex}[0pt][0pt]{% includes a vertical correction for the title page's header
    \shortstack[c]{\sffamily\large Normal, IL \\[2pt] \sffamily\large 2015}%
    \hspace{1.5ex}%
    \IfFileExists{ISUSeal.pdf}{%
      \raisebox{-1ex}{\includegraphics[height=8ex]{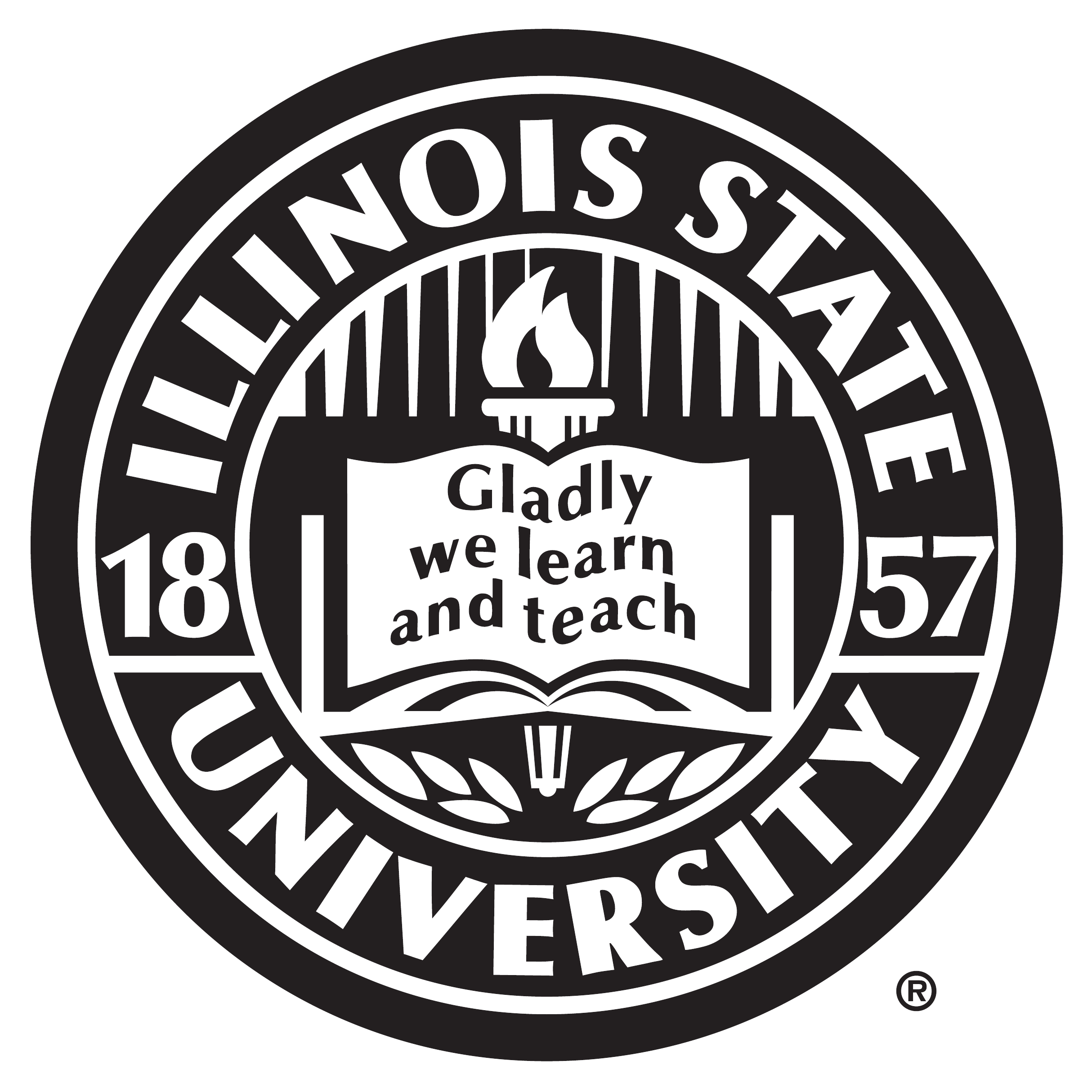}}}{%
      \fbox{\parbox[b][1.5em][c]{5ex}{\tiny Missing picture file!}}%
    }%
  }}

}

\newenvironment{keywords}{%
  \small\vspace{0.5em}\quotation\noindent
  \textbf{\keywordsname:}}
  {\endquotation}
\newcommand\keywordsname{Keywords}
\date{} 
\newcommand\shortauthorlist[1]{\gdef\TheAuthorList{#1}}
  \shortauthorlist{}

\usepackage{amsmath,amssymb,amsthm}

\title{Social Network Analysis of a Grassland Rodent Community Using a Lotka-Volterra Modeling Approach}

\author{
  Majid~ Bani-Yaghoub$^1$, Aaron~ Reed$^2$
  \\[1em]
  $^1$Department of Mathematics and Statistics,\\ University of Missouri-Kansas City,\\ \texttt{baniyaghoubm@umkc.edu}
  \\[1em]
  $^2$School of Biological Sciences,\\ University of Missouri-Kansas City,\\ \texttt{ReedAW@umkc.edu}
  }

\shortauthorlist{Bani-Yaghoub, Reed}

\begin{document}

\maketitle
%%  No indent for Abstract
\begin{abstract}\noindent
Although social network analysis is a promising tool to study the structure and dynamics of wildlife communities, the current methods require costly and detailed network data, which often are not available over long time periods (e.g. decades). The present work aims to resolve this issue by developing a new methodology that requires much less detailed data and it relies on well-established Lotka-Volterra models. Using the long-term abundance data (1973-2003) of northeastern Kansas rodents,  the changes in the magnitude and direction of interactions (e.g., changes from cooperative behavior to competitive behavior or changes in the magnitude of competition) are quantified.
\end{abstract}


%%%%%%%%%%%%%%%%%%%%%%%%%%%%%%%%%%%%%%%%
%%%%%%%%%%%%%%%%%%%%%%%%%%%%%%%%%%%%%%%%
\section{Introduction}

Measuring the long-term changes in population interactions can provide valuable information about the ecology and evolution of wildlife communities. These measures can explain, for example, the coexistence of superior and inferior species in a network of competitors \cite{AllesinaLevine2011}, evolution of competitive interactions \cite{BowlesChoiHopfensitz2003, WolfBrodieMoore1999} and co-evolution of cooperative interactions \cite{Bronstein2009}.  The social network analysis has been frequently used to unpack complex network relationships and to measure the strength links of population interactions \cite{Bascompte2009, Krause2009}. Despite being a promising tool, the network analysis often requires fully detailed data (e.g., GPS data, radio-tracking and capture–mark–recapture data), which is not available for long periods of time (e.g., decades).  Lack of long-term detailed data is a major issue to measure population interactions via network analysis.  A potential solution to this problem is the use of Lotka-Volterra (LV) models, which require much less detailed data (e.g.,  the abundance data of interacting species) to specify the wildlife networks and detect the variations. This approach has the potential to elucidate the presence, magnitude, and variability of competition and other modes of interactions such as facilitation and commensalism.  Recent studies \cite{Geijzendorffer2011, Wu2012} have used empirical data to estimate and monitor the temporal variations in the parameter values of LV models.  Nevertheless, these studies employ indirect linear methods, which can significantly impact the accuracy of model fitting. For instance, the gray LV modeling approach proposed in \cite{Wu2012} relies on a linear programming technique for parameter estimation and model selection. The present work develops a new methodology that will detect and characterize the temporal cycles of population interactions by direct LV model selection and specification. 
\section{Method}
The description of data is available in \cite{BradySlade2004, Reed08}. The competitive LV model of five rodent species residing in northeastern Kansas is given by  
\begin{equation} \label{LV}
\displaystyle \frac{dy_i(t)}{dt} = y_i(t)(a_i-b_iy_1(t)-c_iy_2(t)-d_iy_3(t)-e_iy_4(t)-f_iy_5(t)),
\end{equation}
where all parameters $a_i, b_i, \ldots$ are positive. Table 1 describes the parameters and variables of model (\ref{LV}) and Figure 1 is a schematic representation of the LV network of rodents. The arrows represent the direction of the interactions and the nodes are the populations of the species.
\begin{figure}[h]
		\centerline{\includegraphics[width=1.\textwidth]{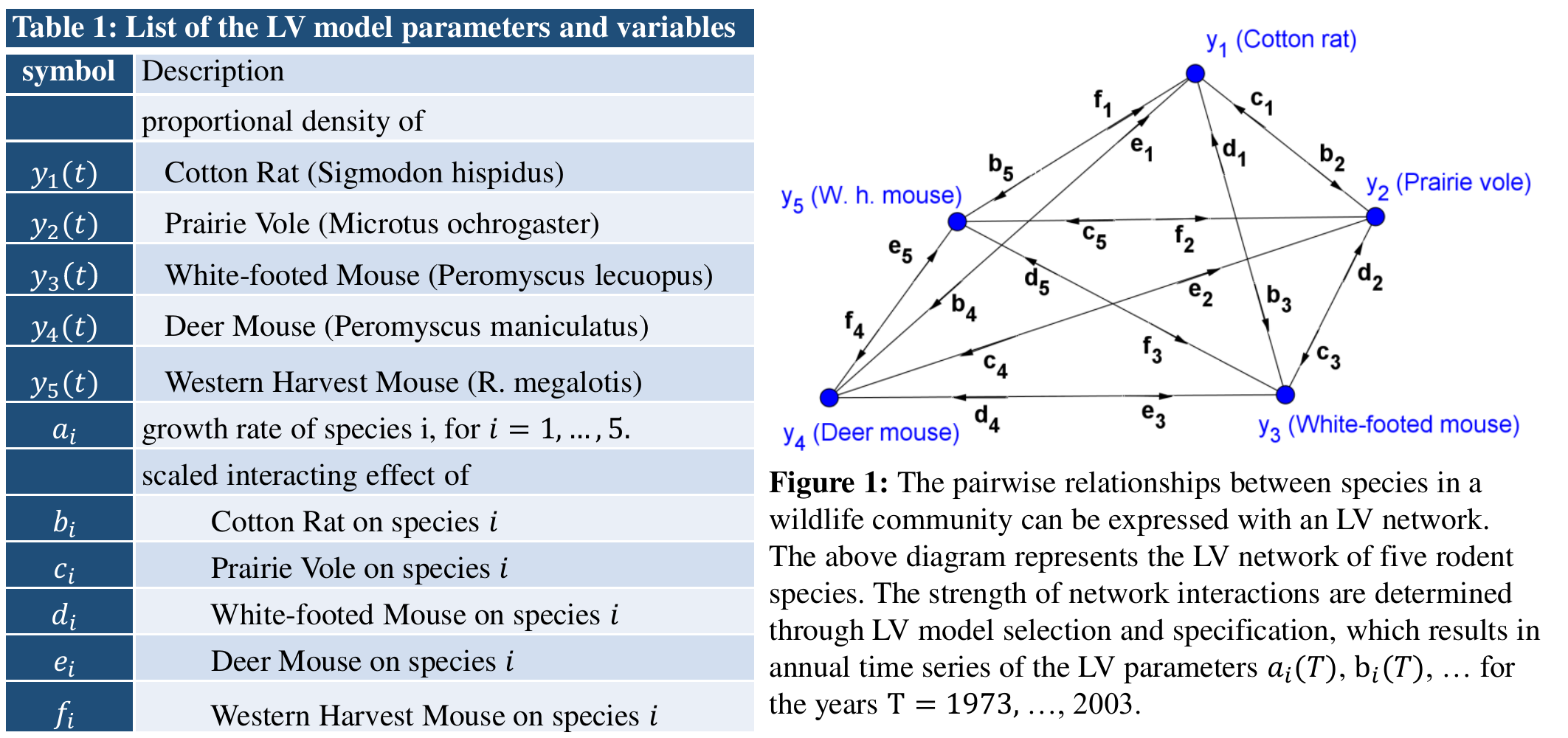}}
		\label{fig:T1}
\end{figure}
%\subsection{Sensitivity Analysis}
Using Matlab the annual competitive LV models were fitted to the long-term abundance data of five Kansas rodent species (see Table 1 for the list of the species). The assumption of annual variations is justified by the analyzing the time series of the abundance data, which indicates the presence of annual cycles in all five species \cite{BradySlade2004, Reed08}. 

\section{Results and Future Work}
 The goodness of fit of the annual competitive LV models is summarized in Table 2, which indicates that the annual LV models are superior to the Sum of Sine Models. Similarly, they outperform various regression models. By analyzing the time series of the estimated parameter values, the level of annually intraspecific competition is measured. Nevertheless, the competition assumption of may impose several restrictions on the parameter estimates and the main outcomes. This assumption is relaxed to include all possible modes of population interactions. The  95\% confidence intervals of the estimated parameter values are obtained though the local sensitivity analysis, which are available  \href{http://b.web.umkc.edu/baniyaghoubm/UMRBAppendixA.xls}{here}. 
 \begin{figure}[h]
		\centerline{\includegraphics[width=1.\textwidth]{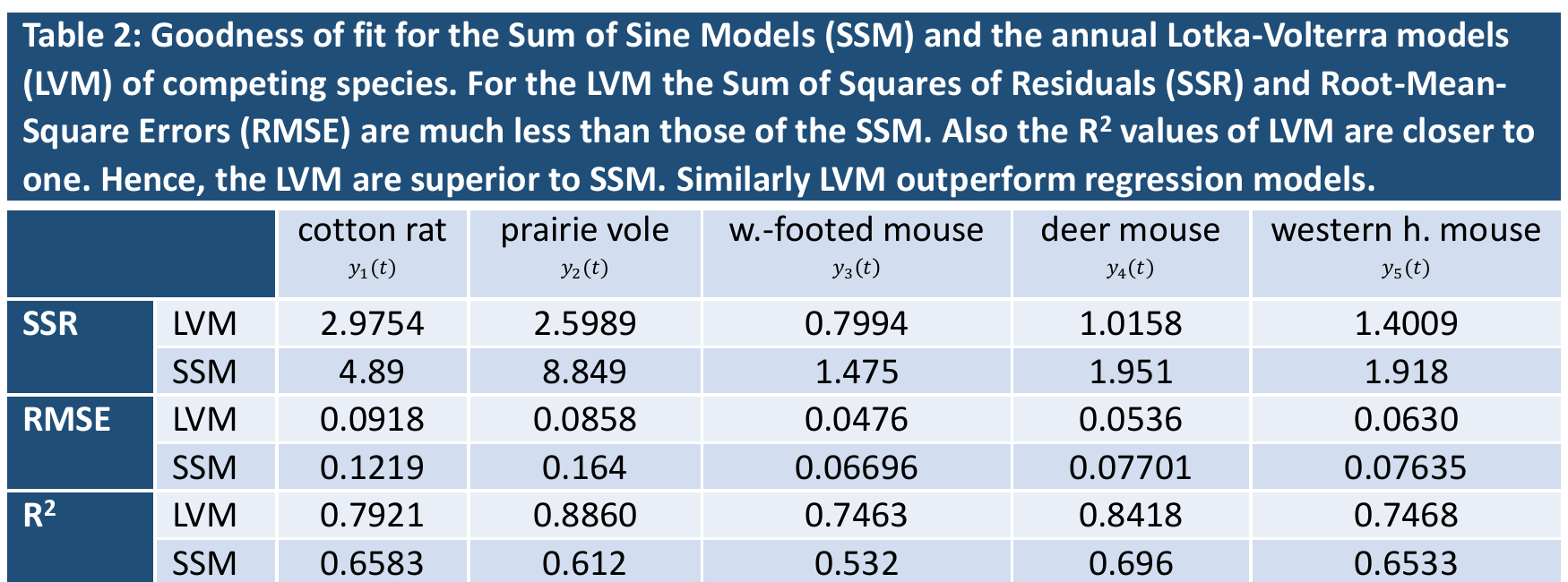}}
		\label{fig:T2}
\end{figure} 
  %\section{Discussion}
Owing to the increasing computational power and the flexibility of LV models, the present project is the first step towards developing a methodology that uses less detailed (but available) data to measure the long-term changes in population interactions. The methodology developed in this study is not limited to the rodent species and it can be employed to unpack the evolution of various wildlife social networks.\\
\indent Model (\ref{LV}) can be extended to a reaction-diffusion model with nonlocality and delay \cite{Bani15, Bani142}. Using the grid data 1973-2003 \cite{Reed08}, the extended model can be used to investigate both spatial and temporal evolution of wildlife social networks. The extended model is validated using the pseudo-independent data set (data collected from 2007-2014 from the same grid but not included in the above analyses). This will constitute the further development of a methodology to analyze the spatio-temporal evolution of population interactions in various wildlife communities, and develop a framework to study the possible impacts of climate change on the evolution of wildlife social networks.  %Given that our long-term dataset includes both space and time trapping record of the rodents, the proposed methodology can be validated both with respect to time and space.  
%%%%%%%%%%%%%%%%%%%%%%%%%%%%%%%%%%%%%%%%
%%%%%%%%%%%%%%%%%%%%%%%%%%%%%%%%%%%%%%%%
%\appendix
%\section{Appendix A}
%The estimated values are available 
%%%%%%%%%%%%%%%%%%%%%%%%%%%%%%%%%%%%%%%%
%%%%%%%%%%%%%%%%%%%%%%%%%%%%%%%%%%%%%%%%

\end{document}